\documentstyle[12pt]{article}
\renewcommand{\theequation}{\thesection.\arabic{equation}}
\newcommand{\be}{\begin{equation}}
\newcommand{\ee}{\end{equation}}
\newcommand{\bear}{\begin{eqnarray}}
\newcommand{\eear}{\end{eqnarray}}
\newcommand{\ba}{\begin{array}}
\newcommand{\ea}{\end{array}}

\newskip\humongous \humongous=0pt plus 1000pt minus 1000pt

\newif\ifdtup

\def\oldreffmt#1{\rlap{[#1]} \hbox to 2\parindent{}}

\def\figfmt#1{\rlap{Figure {#1}} \hbox to 1in{}}

\def\ie{\hbox{\it i.e.}{}}	
\def\eg{\hbox{\it e.g.}{}}

\def\Tr{\mathop{\rm Tr}}

\def\bra#1{\left\langle #1\right|}
\def\ket#1{\left| #1\right\rangle}

\def\slash#1{#1\!\!\!/\!\,\,}
\def\beq{\begin{equation}}
\def\eeq{\end{equation}}
\def\bea{\begin{eqnarray}}
\def\eea{\end{eqnarray}}
\def\half{\frac{1}{2}}

\def\bq{\begin{quote}}
\def\eq{\end{quote}}

\def\half{\frac{1}{2}}     

\relax
\newdimen\tdim
\tdim=\unitlength
\def\bar{\overline}

\begin{document}

\pagestyle{empty}
\begin{titlepage}
\def\thepage {}    

\title{\bf Dirac Branes and\\
Anomalies/Chern-Simons terms in any $D$ 
 \\ [2mm] } 
\author{
\bf  Christopher T. Hill \\[2mm]
 {\small {\it $^1$Fermi National Accelerator Laboratory}}\\
{\small {\it P.O. Box 500, Batavia, Illinois 60510, USA}}
\thanks{e-mail: hill@fnal.gov}
\\ [3mm]
}

\date{ }
\maketitle

\noindent

\vspace*{-9.0cm}
\begin{flushright}
FERMILAB-Pub-09/282-T \\ [1mm]
June 19, 2009 
\end{flushright}

\vspace*{8.1cm}
\baselineskip=18pt

\begin{abstract}
The Dirac quantization procedure of a magnetic
monopole can be used to derive the
coefficient of the $D=3$  Chern-Simons term through a self-consistency argument,
which can be readily generalized to any odd $D$. This yields consistent and
covariant axial anomaly coefficients on a $D-1$ boundary,
and Chern-Simons term
coefficients in $D$.  In $D=3$ 
magnetic monopoles
{\em cannot exist} if the Chern-Simons $AdA$ term is present.
The Dirac solenoid then becomes a physical closed 
string carrying electric current.
The charge carriers on the string must be consistent with
the charge used to quantize the Dirac solenoidal flux.
This yields the 
Chern-Simons term coefficient.
In higher odd $D$ the intersection of $(D-1)/2$ Dirac branes
yields a charged world-line permitting the consistency
argument.  The covariant anomaly
coefficients follow readily from generalizing the counterterm. 
This purely bosonic derivation of anomalies is quite simple, 
involving semiclassical evaluation of exact integrals,
like $\int dAdA...dA$, in the brane intersections.
\end{abstract}

\vspace*{1cm}
\vfill
\end{titlepage}

\baselineskip=18pt
\renewcommand{\arraystretch}{1.5}
\pagestyle{plain}
\setcounter{page}{1}


\section{\bf Introduction}

Two of the most fundamental results in quantum field theory are the scale anomaly and axial anomaly. Both effects begin at order $\hbar$ (one-loop), and represent true quantum breaking of classical symmetries.  The scale anomaly is equivalent to the 
renormalization group $\beta$-function and running of the coupling constant and through it quantum field theory can establish a fundamental mass scale
by dimensional transmutation, as happens in the case of QCD (for
a review see \cite{cth0} and references
therein).

The axial anomaly  \cite{S,BJ,Adler,Bardeen,AdlerBardeen}
arises in even $\hat{D}$ (generally we'll use $\hat{}$ to denote even
dimension). Weyl spinor loops avoid many ambiguities 
and yield the 
``consistent anomaly,'' which has the form in a $U(1)$ theory,
$(e/2\pi)^{\hat{D}/2}/ ((\hat{D}+2)/2)!) \epsilon_{\mu\nu ...\rho\sigma}
\partial^\mu A^\nu...\partial^\rho A^\sigma $. Fermion 
loop calculations that yield this simple result, however, are 
rather cumbersome \cite{framp}.
This result is equivalent to the Chern-Simons (CS) term coefficient
one dimension higher, since the $\hat{D}+1$ CS term generates the anomaly
on a boundary under a gauge transformation. There are various arguments to
fix the coefficients of the CS term, often in the context
of normalizing the charges of solitons arising in the nonabelian case. 
We presently seek a transparent argument for the general case within
a $U(1)$ gauge theory. 

We will give presently a simple
illustration as to how the axial anomaly arises directly from 
Dirac's construction of the magnetic monopole in $D=3$
for pure electrodynamics.  First we note
that Dirac monopoles {\em do not exist when the Chern-Simons term is present}.  
The conserved Chern-Simons current requires that
Dirac solenoids, which carry quantized magnetic flux,
must then become closed loops.  These loops carry
electric current, or in $D=1+2$, the solenoids become the world-lines of charged particles.
The Dirac solenoid in any $D$-odd
generalizes to a $D-2$ dimensional hypersurface, or Dirac brane, first
considered by Teitelboim \cite{teit}, 
to which a quantized electromagnetic flux is attached.  The intersection
of $(D-1)/2$  Dirac branes becomes a charged
particle world-line when the Chern-Simons term
is turned on. 

Our essential trick, therefore, is to note that in
any $D$-odd there is always a configuration of electromagnetic
fields (not necessarily a solution to
equations of motion), typically an intersection of Dirac branes, that
forms a charged particle world-line in the presence of the CS term. 
The resulting electric charge of this special
configuration must then be consistently set equal to the original charge
used to quantize the Dirac flux.  This ``bootstrapping'' 
condition then dictates the coefficient of the CS term in any odd $D$. The
result depends only upon 
exact integrals, \eg, the ``core structure'' of a Dirac brane is irrelevant. 
The boundary of the CS term under a gauge transformation
yields the consistent anomaly. 
We furthermore obtain the ``covariant anomaly'' 
by generalizing the Adler-Bardeen counterterm.  

This analysis is carried out in a
$U(1)$ theory, but the result
is general to any nonabelian theory.  The result is completely bosonic in origin, requiring no fermion loop calculation. It virtually
reduces to a ``back of an envelope'' computation.
In nonabelian theories the CS term controls the properties of
various solitons, such as instantonic vortices in $D=5$.
To us, the present result illuminates why the $D$-odd CS term in a $U(1)$ theory, $
\epsilon_{\lambda\mu\nu ...\rho\sigma}
A^\lambda\partial^\mu A^\nu...\partial^\rho A^\sigma$,  exists at all 
and what it is physically measuring, \ie, the world-line intersection of Dirac branes.

\section{Dirac Monopoles and the Chern-Simons Term}

\subsection{The Dirac Monopole}

The intertwining of quantum physics 
and topology begins with Dirac's construction of the
magnetic monopole in $D=3$ (see reviews of \cite{Nash}, \cite{Polchinski}).
We give a quick review 
in this section.

Dirac imagined an idealized solenoid in
three space dimensions carrying a magnetic flux
$\Phi = \int_S B\cdot d(area)$ where the integral
is over a
cross-section of the solenoid.
The mass per length of the solenoid is neglected.
The solenoid can be viewed as an infinitely long ray
terminating at a 
point in space, $\vec{x}_0$.
At $\vec{x}_0$ the magnetic flux emerges from the open
end of the solenoid.
The magnetic charge of the monopole is given
by Gauss' law:
\beq
4\pi g_m = \Phi
\eeq

\begin{figure}[t]  
\vspace{4.5cm}  
\includegraphics{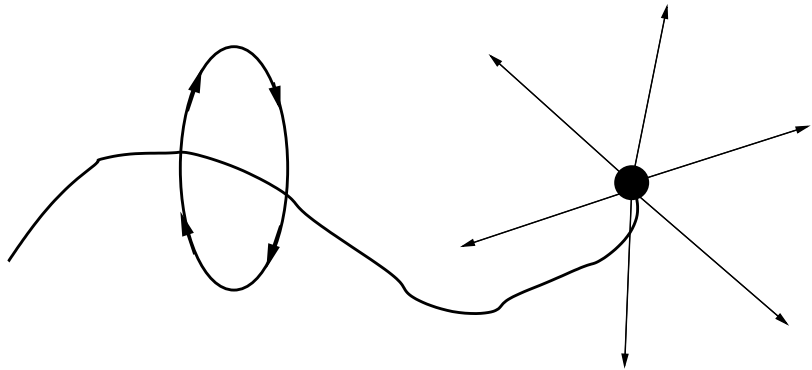}  
\vspace{3.5cm}  
\caption[]{
\small 
\addtolength{\baselineskip}{-.3\baselineskip}  
 Dirac monopole construction with solenoid. The
flux is quantized such that the Aharonov-Bohm phase for
an encircling electron is a multiple of $2\pi$. }  
\label{dirac}  
\end{figure}  

Dirac asked how to make the solenoid undetectable in electrodynamics?
Classically one can hide the solenoid arbitrarily
well by making it have an arbitrarily small cross-sectional area.
However, the infinitesimal solenoid remains detectable
at large distances in quantum mechanics. External to the solenoid there is a circumferential vector potential.
By Stoke's theorem:
\beq
\Phi = \oint \vec{A} \cdot d\vec{z} 
\eeq
The line integral over $\vec{A}$ determines the phase shift,
$\phi$, of an electron
wave-function  as the electron, of charge $e$, loops the solenoid
(the Aharonov-Bohm phase):
\beq
\phi = e\oint \vec{A} \cdot d\vec{z} /\hbar
\eeq
For arbitrary $\phi$ the solenoid is observable
as a diffraction pattern of an impinging, long wave-length electron beam.
However, if $\phi= 2\pi N$ then this phase shift is unobservable
and the solenoid is ``cloaked.''
This implies that the magnetic flux must be quantized 
as $\Phi =\hbar\phi/e = 2\pi\hbar N/e$.
Thus the magnetic charge of a Dirac monopole is then quantized:
\beq
g_m = \frac{N\hbar}{2e}
\eeq
In the following we will consider the case $N=1$.

The solenoid part of the construction can be removed 
by describing a single monopole
with two vector potentials, respectively on
the left (right) hemisphere
of a monopole with Dirac solenoid running in from the right (left).
Wu and Yang  \cite{Wu}  demand that these two potentials  are gauge equivalent
in an overlapping region.  This latter 
consistency condition then enforces the
monopole quantization condition. In this sense the
solenoidal becomes an artificial component of the
construction.  

Our present perspective, however, is exactly the opposite: we
 keep the solenoid in $D=3$, and in fact,
we  {\em discard the monopole}.
In fact, {\em we must discard the monopole} when the CS term
is incorporated into the theory, as we will now discuss.
The fate of the solenoid then becomes interesting:
The Dirac solenoids can be taken to be  
closed loops of string, which resemble closed bosonic strings, or
infinite time-like world-lines in $D=1+2$ Minkowski space.

\subsection{D=3 Electrodynamics with Chern-Simons Term}

Let us now incorporate the CS term into
the action of $D=3$ electrodynamics:
\beq
S_{CS} = -\kappa \; \int d^3 x \; \epsilon_{\mu\nu\rho}A^\mu \partial^\nu A^\rho =  -\kappa\int AdA =  -\kappa\int \vec{A}\cdot(\vec{\nabla}\times \vec{A}) 
\eeq
The CS term depends explicitly upon the vector
potential and forces the parameterization of the electromagnetic
field, $F_{\mu\nu}$, to be determined by it. We
also include the kinetic term $(-1/4)\int \; F_{\mu\nu}F^{\mu\nu} \rightarrow
(1/2)\int \vec{A}\cdot(\nabla^2\vec{A} -\vec{\nabla}(\vec{\nabla}\cdot\vec{A})) $
into the action.

The resulting Maxwell's
equations are modified by the presence of the CS term,
\beq
\label{maxwell}
\nabla^2 \vec{A} -\vec{\nabla}( \vec{\nabla}\cdot \vec{A}) = 2 \kappa \vec{B}
\eeq
where $\vec{B} = \vec{\nabla}\times \vec{A}$.
Crossing the Maxwell equation
with $\vec{\nabla}$ we have,
$\nabla^2 (\vec{\nabla}\times \vec{A}) = 2 \kappa \vec{\nabla}\times (\vec{\nabla}\times \vec{A} )
= 2 \kappa(\nabla^2 \vec{A} -\vec{\nabla}(\vec{\nabla}\cdot \vec{A})) $
and, using the Maxwell equation again on the {\em rhs}, we thus have:
\beq
\label{B}
\nabla^2 \vec{B} = 4 \kappa^2  \vec{B}
\eeq
This latter form displays the fact that
the CS term induces a mass for the photon \cite{schonfeld,deser}
(in $D=5$ or higher the CS term induces interactions amongst KK-modes 
that typically violate T-parities \cite{cth3}).

However, with the magnetic field
defined as usual, $\vec{B} = \vec{\nabla} \times \vec{A}$,
we have everywhere outside the solenoid:
\beq
\label{2.8}
\vec{\nabla}\cdot \vec{B} = 0
\eeq
From the equation of motion eq.(\ref{B}) a Dirac
monopole, with nonzero $\kappa$,
would have to produce a radial magnetic field that
attenuates in the Yukawa form: $\vec{B} \propto \vec{\nabla}\phi$
where $\phi = (\exp(-2 \kappa r)/r)$.
This would require a corresponding nonzero $\nabla\cdot B \propto \vec{\nabla}^2 \phi
= 4\kappa^2 \phi $ everywhere, violating eq.(\ref{2.8}). 
Hence, a radial magnetic field, or a corresponding
magnetic charge cannot exist with $\kappa \neq 0$ !
The CS term requires that solenoids not
terminate with open ends. Solenoids thus
become closed loops carrying electric current.

\begin{figure}[t]  
\vspace{4.5cm}  
\includegraphics{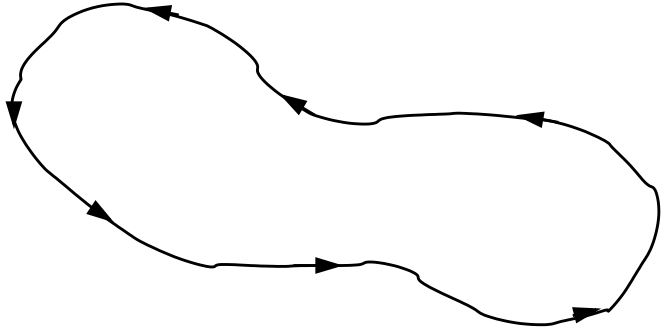}  
\vspace{3.5cm}  
\caption[]{
\small 
\addtolength{\baselineskip}{-.3\baselineskip}  
The solenoid loop becomes a current loop when the CS
term is included into the action. Demanding that
the current carrier charge is $e$ fixes
the coefficient of the CS term: $\kappa =e^2/4\pi$.
}  
\label{solenoid}  
\end{figure}

Consider the Dirac solenoid loop in Fig.(2).
Let us describe the loop by coordinate $\vec{x}(\tau)$
parameterized by $\tau$.  The solenoid has a circumferential
vector potential, $\vec{A}_{solenoid}$, and a
``core'' field $\vec{B}_{core} = \vec{\nabla}\times \vec{A}_{solenoid}$.
$\vec{B}_{core}$ is a singular field attached to the solenoid loop.
Consider the coupling to an external ``photon,'' 
$ \vec{A}^{ext}$,  by the solenoid loop.
Here we effectively evaluate the matrix element
of $AdA=\vec{A}\cdot (\vec{\nabla}\times \vec{A}) = \vec{A}\cdot \vec{B}$ 
in a coherent state containing the solenoidal field
and external field: $\ket{\vec{B}_{core}, \vec{A}^{ext}}$.
The matrix element $\bra{\vec{B}_{core}, \vec{A}^{ext}} AdA \ket{\vec{B}_{core}, \vec{A}^{ext}}$
becomes $2\vec{A}^{ext}\cdot \vec{B}_{core}$, where the factor of $2$ arises 
because we can contract the 
fields in $AdA$ with the internal flux and external photon in two
ways (see discussion in section 3.1 on coherent states).

If we integrate over volume, we integrate out
the transverse dimensions (the solenoid cross-section),
using $\int d(area) \vec{B}_{core} = ({2\pi \hbar}/{e}) d\vec{x}(\tau)/d\tau $,
and
the CS term takes the form:
\beq
-\frac{4\pi  \kappa}{e} \; \int d\tau \; \; {A}^{ext}_\mu \frac{dx^\mu}{d\tau}
\eeq
This is just the action of a charged classical current
loop, carrying charge $q=4\pi  \kappa/e$.
The details of the singular core magnetic field have
 disappeared by integrating over the transverse 
 dimensions of the solenoid
 and the result depends only upon the
 quantized flux $\Phi$. 
 We can apply this result to $1+2$ dimensions where the solenoid string
is stretched out to become a timelike world-line (returning
to the past as an antiparticle). The CS term 
on the world-line becomes:
\beq
 -\kappa\; \int d^3 x \; \bra{\vec{B}_{core}, \vec{A}^{ext}}\epsilon_{\mu\nu\rho}A^\mu \partial^\nu A^\rho
\ket{\vec{B}_{core}, \vec{A}^{ext}} \rightarrow
-\frac{4\pi  \kappa}{e} \int dx^\mu {A}^{ext}_\mu
\eeq
Consistency demands that we equate the induced electric charge $q$ of the
solenoid world-line to the same value, $e$, that is the defining
charge of Dirac's quantization condition.
Hence we must have:
\beq
\label{coeff}
 q= e\qquad \qquad \makebox{or,} \qquad \kappa = \frac{e^2}{4\pi}
\eeq
This bootstrapping condition
fixes the value of the CS term coefficient, $\kappa$. 

Note that the CS current is obtained from the action
by varying {\em wrt}  $A_\mu$:
\beq
J_\mu = -\frac{\delta}{\delta A_\mu}S_{CS} 
=  2\kappa\; \epsilon_{\mu\nu\rho}\partial^\nu A^\rho
\eeq
and we see that automatically:
\beq
\partial_\mu J^\mu = 0
\eeq
From eq.(\ref{maxwell}) we see that the conserved CS current
is the source term for the modified
Maxwell's equation. Of course,  the CS
current is just the magnetic field and the conservation
law is just $\vec{\nabla}\cdot \vec{B} = 0$.  
The conservation law of the CS current is 
forcing the solenoid to be a closed loop with the
conservation of electric charge. 

Another way of getting
the quantization
of the solenoid charge is to
note that solenoid strings have nontrivial Gauss-linking.  If we link two
closed loops we can view one loop as the test particle used to measure the flux
in the other.  The result involves the contributions
to each loop from the other and thus takes the form:
\bea
 \kappa\; \int d^3 x \; \epsilon_{\mu\nu\rho}A^\mu \partial^\nu A^\rho &  \rightarrow & 
 \kappa\int A_1dA_2 + A_2 dA_1 
\nonumber \\
& = &  2 \kappa\times \left(\frac{2\pi}{e}\right)^2
\nonumber \\
& = &  2\pi
\eea
Thus, the CS term with the Dirac quantization
condition implies that the action shifts by (a multiple
of) $2\pi$ under linking.  This enforces locality of the 
interactions of the strings. That is, the only observable
consequence of linking a pair of loops would
be seen at the point of intersection, and not in a nonlocal
overall arbitrary phase shift of the path integral.
Gauss linking is another way to argue for the consistency
condition for the Chern-Simons term coefficient
(see an alternative recent derivation of Witten, \cite{wit2}).

\begin{figure}[t]  
\vspace{4.5cm}  
\includegraphics{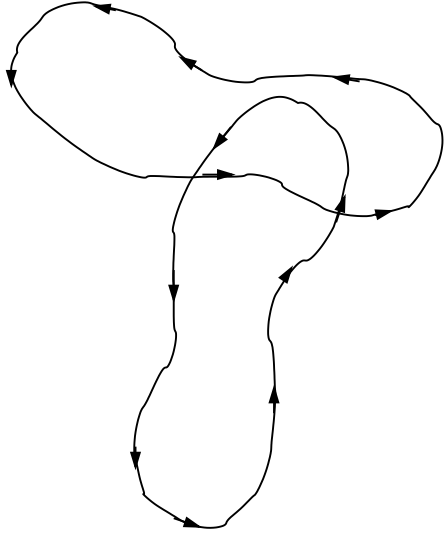}  
\vspace{3.5cm}  
\caption[]{
\small 
\addtolength{\baselineskip}{-.3\baselineskip}  
Two solenoid loops are linked
and the action shifts by a multiple of $2\pi$
with the quantized CS term coefficient.
}  
\label{solenoid}  
\end{figure}

We see, however, that if two stationary solenoidal world-lines  
are exchanged in spatial position,  the corresponding phase shift
 is exactly half of a Gauss linking of two loops. The action
must therefore shift by $\pi$. 
Ergo, Dirac world-line solenoids are fermions. 
If we relaxed the consistency
condition that enforces 
$ \kappa = e^2/4\pi$, then the exchange of solenoid particles 
leads to an arbitrary shift in the action.  The particles become
``anyons'' with arbitrary statistics \cite{wil}. 
From another perspective, if we have  electrons that already
have spinor (fermionic) statistics,
and we turn on the CS
term, the electrons can become bosons. 
This raises various intriguing possibilities, \eg, that
electrons in a  $1+2$ dimensional systems could undergo Bose condensation
without Cooper pairing. The interplay of the CS term  suggests
various components
of a model of high-$T_c$ superconductivity.

Note that the profile of the 
Dirac solenoid on a $\hat{D}=2$ spacelike hypersurface is 
the current loop containing $\epsilon_{ij}F^{ij}$, \ie, it is
an instanton. Instantons in even $\hat{D}$
can be viewed as events in which solenoids (or instanton vortices in nonabelian models)
in $\hat{D}+1$ are created or destroyed on boundary
branes.   The discussion of instantons
is beyond our present scope, involving the exchange of solenoids
(instantonic vortices) between boundary branes containing 
chiral fermions.
However,  this construction
leads to the correct normalization of the instanton,
\ie, the Pontryagin index, and we can show that this is related
directly to the covariant anomaly.

As an aside, this raises the question of what the solenoid core structure
is? We emphasize that nothing in our present results depend upon the core.
In a Nielsen-Olesen flux-tube vortex, the Higgs
field develops
a nontrivial radial profile that approaches 
a constant as one moves away from the vortex, and
carries a circulating current density. The vacuum thus has
a distributed charge and current density which supports the 
core magnetic field structure, and the magnetic field is confined
in the core with a radial profile. The Higgs field at infinity
carries a phase factor that is proportional to the azimuthal angle
and thus maps the $U(1)$ gauge group onto the circle at infinity.
In this way the Nielsen-Olesen flux tube is an element of the homotopy
group $\pi_1(U(1))$ which is the topological definition of a vortex.
In the present case, the loop of the gauge field, $\epsilon_{ijk}
\partial^j \vec{A}^k $
is promoted to a physical circulating current by the CS term.
Thus the $U(1)$ gauge group is mapped onto the circle at
infinity by the Wilson line containing the 
circulating $\vec{A}$. In this way the Dirac solenoid is
an element of the homotopy group $\pi_1(U(1))$, and behaves as a vortex.
In the case of the Dirac solenoid, however, the core has collapsed.
Indeed,
the core singularity is not a solution of  eq.(\ref{B}) either, and
we must invoke new short-distance physics to support the core flux. 
This is analogous to the Skyrmion in the absence of the Skyrme term.
The Skyrmion is on a $D=3$ space-like hypersurface, defined by
$U=\exp(if(r)\hat{r}\cdot \vec{\tau}/2)$ 
with a radial core profile such that $f(0) = 0$ and $f(\infty) = 2\pi$.
It thus is both an element of $\pi_3(SU(2))$, mapping the $SU(2)$
gauge group onto the full manifold,   and it is
also an element of $\pi_2(SU(2))$, mapping $SU(2)$ onto the surface $S_2$
at infinity. When the core of the Skyrmion collapses, $f(r) \propto \theta(r)$,
it ceases to be
an element of $\pi_3(SU(2))$, but remains an element of $\pi_2(SU(2))$.
Since our discussion is mainly focused on the large distance topological
aspects of the solenoid, we will presently sidestep the issue of the Dirac solenoid core structure, and assume that we can model it by suitable
extensions of the theory (\eg, in analogy to adding the Skyrme term; perhaps
gravity is involved).

\subsection{Fermionic Anomalies on Boundaries}

This section deals with compactification of $D=3$ and 
chiral delocalization of fermions on the boundary
of $\hat{D}=2$ (the $D=5$ to $\hat{D}=4$ case is treated
in ref.(\cite{cth3})), and it can be skipped
on a first reading. The purpose is to illustrate the interplay of
fermionic anomalies with the CS term to maintain
an overall gauge invariant theory. The CS term
provides a counterterm which converts the axial anomaly
into the ``covariant'' form automatically.
We return to the mainline discussion of
Dirac brane arguments in higher $D$ in section 3.

We can compactify $D=3$ to $\hat{D}=2$
by sandwiching pure $D=3$ QED between branes
separated by $R$
(\eg, ``capacitor plates,'' in an older lexicon).
The vector potential exists in the bulk, and
performing a local gauge transformation, 
$\delta {A}_\mu = -{\partial_\mu}\theta$,
we see that the CS term generates a surface term on the 
branes (for convenience, we set $e=1$):
\beq 
\left.\delta S_{CS} =  \int d^2x\;
\frac{\theta(x_\mu,x^3) }{4\pi}\epsilon_{\mu\nu}\partial^{\mu}A^{\nu}(x^\mu,x^3)\right|_{x^3=0}^{x^3=R}.
\eeq
The CS term shifts the action on the boundaries
and thus yields anomalies on the $\hat{D}=2$ branes.

\begin{figure}[t]  
\vspace{4.5cm}  
\includegraphics{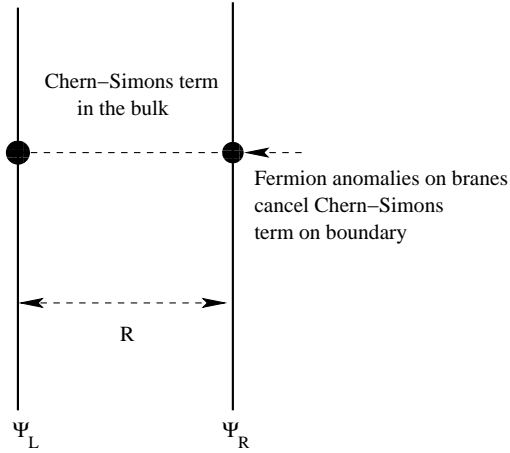}  
\vspace{2.5cm}  
\caption[]{
\small 
\addtolength{\baselineskip}{-.3\baselineskip}  
Orbifold with split, anomalous fermions (electrons).
$\psi_L$ ($\psi_R$) is attached to the $\hat{D}=2$ left-brane, (right-brane).
Gauge fields propagate in the $D=3$ bulk, which has a compactification
scale $R$. The bulk contains a CS term, and the branes
produce loop diagram amplitudes in the effective action. The anomalies
from the CS term cancel the anomalies from the triangle diagrams
on the respective branes so the overall theory is anomaly free. }  
\label{dirac2}  
\end{figure}

We introduce chiral fermions that are constrained to the 
boundary branes: 
\bea
& & \int d^2x\; \bar{\psi}_L(i\slash{\partial} - \slash{A}_L)\psi_L
\qquad \makebox{brane $I$}
\nonumber \\
& & \int d^2x\; \bar{\psi}_R(i\slash{\partial} - \slash{A}_R)\psi_R
\qquad \makebox{brane $II$}
\eea
where chiral projections are $\psi_{L,R} = \half(1\mp\gamma^3)$.
Here we have the gauge fields on the boundary branes:
\be
A_L(x_\mu) = A(x_1,x_2,x^3=0)\qquad  A_R(x_\mu) = A(x_1,x_2,x^3=R)
\ee
The anomaly is
readily computed for massless Weyl fermions in $\hat{D}=2$ by performing
the one-loop current bubble with an external gauge field (see the Appendix):
\beq
\label{fff}
\partial_\mu j^\mu_L = -\frac{1}{4\pi} \epsilon_{\mu\nu}\partial^\mu A_L^\nu
\qquad\qquad
\partial_\mu j^\mu_R = \frac{1}{4\pi} \epsilon_{\mu\nu}\partial^\mu A_R^\nu
\eeq
(recall, $j_{L,R} = -\delta S/\delta A_{L,R} = \bar\psi\gamma_\mu(1\mp\gamma^3)\psi/2 $).

Classically the theory is invariant if the fermions
transform on the boundary
as $\psi_L\rightarrow e^{i\theta(0)}\psi_L$ and  
$\psi_R\rightarrow e^{i\theta(R)}\psi_R$, when 
in the bulk, $\delta A_\mu = -\partial_\mu \theta $.
However, the action on
the branes also shifts due to the fermionic anomalies, under 
 $\delta A_\mu = -\partial_\mu \theta $, as:
\bea 
\delta S_{\delta A} & =&  -\int d^2x\; \theta(0)\partial^\mu \bar{\psi}_L\gamma_\mu\psi_L
-\int d^2x\; \theta(R)\partial^\mu \bar{\psi}_R\gamma_\mu\psi_R
\nonumber \\
& = & \left. -\int d^2x\;
\frac{\theta(x_\mu,x^3)}{4\pi}\epsilon_{ab}\partial^aA^b\right|_{x^3=0}^{x^3=R}.
\eea
whence:
\be 
\delta S_{CS} + \delta S_{\delta A} = 0
\ee
We have thus arranged a cancellation of
the CS anomalies of eq.(\ref{fff}) by chiral fermions
on the boundaries.  This is a ``chiral delocalization''
in $D=3$ leading to a compactified theory with a Dirac spinor
in $\hat{D}=2$.\footnote{Note that we can give the $D=3$ fermions a mass by way of a
bilocal operator $m\bar{\psi}_L(0)W\psi_R(R) $ where
$W$ is a Wilson line connecting the branes (see \cite{cth3})}.

It is convenient to write the anomalies in the ``$VA$'' form
using: 
\beq
\label{VA}
A_L = V-A,\qquad  A_R = V+A,\qquad j_V=j_R+j_L,\qquad
j_A=j_R-j_L
\eeq
whence:
\beq
\partial_\mu j^\mu_V = \frac{1}{2\pi} \epsilon_{\mu\nu}\partial^\mu A^\nu
\qquad\qquad
\partial_\mu j^\mu_A = \frac{1}{2\pi} \epsilon_{\mu\nu}\partial^\mu V^\nu
\eeq
These are called the ``consistent anomalies.'' In a nonabelian
theory they are generally not gauge invariant operators
but satisfy the Wess-Zumino consistency conditions.

In the $D=3$ bulk, let us consider only the three
lowest KK modes, corresponding
to a vector zero mode, and axial vector cosine mode,
and a pseudoscalar, sine mode ($x^\mu = (t, x)$, $x^3 = y$):
\beq
\hat{A}_\mu(x^\mu, y) = V_\mu(x^\mu) - A_\mu(x^\mu)\cos(\pi y/R)
\qquad
\hat{A}_3(x^\mu, y) = \phi \sin(\pi y/R )
\eeq
The sign of $A_\mu$ relative to $V_\mu$
is chosen so that eq.(\ref{VA}) is satisfied for $A_{L,R}$.
Orbifold boundary
conditions equivalently follow from the assumption that
$F_{\mu 3}(x^\mu,y=0) = F_{\mu 3}(x^\mu,y=R) = 0$.

Evaluating the CS term in the
truncated $\hat{A}_\mu(x^\mu, y) $  yields:
\beq
S_{CS} = -\frac{1}{4\pi} 
\int d^2x\; \int dy \;\epsilon_{\mu\nu\rho}\hat{A}^\mu\partial^\nu \hat{A}^\rho
= \frac{1}{2\pi} \int d^2x\; \epsilon_{\mu\nu}V^\mu A^\nu
+\frac{1}{2\pi f} \int d^2x\; \phi \epsilon_{\mu\nu}\partial^\mu V^\nu
\eeq
where $f= \pi/4R$, and note in the first term on the {\em rhs}
a tricky sign coming from $\epsilon_{\mu 3 \nu} = -\epsilon_{\mu \nu}$. 
The ``pion'' $\phi$ couples anomalously
to the vector field, but this simple orbifold model
$\phi$ is eaten by $A$, and we define $\tilde{A} = A + \partial \phi/f$.
The remaining term, $\sim \epsilon_{\mu\nu}V^\mu \tilde{A}^\nu $, acts as a counterterm 
in the action. Its presence 
modifies the currents:
\beq
 \delta j^\mu_V = -\frac{\delta S_{CS}}{\delta V_\mu}
 =
 -\frac{1}{2\pi} \epsilon_{\mu\nu}\tilde{A}^\nu
\qquad
 \delta j^\mu_A = -\frac{\delta S_{CS}}{\delta \tilde{A}_\mu}
 =
 \frac{1}{2\pi} \epsilon_{\mu\nu}V^\nu 
\eeq
and we thus define the full currents, 
\beq
\tilde{j}_V = j_V + \delta j_V
\qquad
\tilde{j}_A = j_A + \delta j_A
\eeq
and we find:
\beq
\partial_\mu \tilde{j}^\mu_V = 0
\qquad
\partial_\mu \tilde{j}^\mu_A = \frac{1}{\pi} \epsilon_{\mu\nu}\partial^\mu V^\nu 
\eeq
These latter forms are the
``covariant" anomalies. The vector current is now conserved
reflecting the overall gauge invariance with the fermions on branes and
CS term in the bulk.  
The corresponding analysis in compactifying $D=5$ QED to
$\hat{D}=4$ QED is given in \cite{cth3} with more detail on
Wilson line fermion masses and the full KK-tower anomaly structure.
The main lesson is that the $D=\hat{D}+1$ CS term becomes the $\hat{D}$ counterterm, and
it generally impacts the physics of the $\hat{D}$ theory.

\section{Generalization to Dirac Branes in any odd D}

To construct a solenoid in $D=3$ we 
effectively ``stack'' $(xy)$
plaquette current loops along the $z$ axis.  The loop integral of $A_\mu$ in
the $(xy)$ plane circumnavigating the stack
is then Dirac quantized. By Stoke's theorem, the surface
integral spanning the loop, hence the solenoidal flux, is likewise 
Dirac quantized.

To generalize this construction, let us first consider 
the special case of $D=5$. 
An $(xy)$ plaquette bounded by a current loop
can be ``stacked''  simultaneously along
all of the three orthogonal axes, $(zwt)$.  We thus start with one 
``kernel'' $(xy)$
plaquette and we stack with three new $(xy)$ plaquettes taking
infinitesimal steps in the $z$, $w$, and $t$ directions
respectively.  We then iterate the procedure, generating nine more plaquettes,
etc.  Therefore, the resulting stack of current
plaquettes spans a $3$-dimensional hypersurface. This hypersurface
carries the field strength dual to $F_{xy}$, \ie,
$F^*_{zwt}= F_{xy}$.  This is a Dirac brane flux in the $zwt$ hypersurface.

In the $xy$ plane the resulting 
$3$-dimensional hypersurface can be  encircled in $D=5$ by a closed
loop and the flux in the hypersurface, $F^*_{zwt}$,
can be Dirac quantized. We have, integrating over a loop in the $(xy)$ plane
(bounding a disk):
\beq
\label{fid}
\oint dx^\mu A_\mu = \int_{disk} dx dy \;F_{xy}  
= \int_{disk} dx dy \;F^*_{zwt} = \frac{2\pi}{e}
\eeq
This gives the generalization of the cloaked Dirac solenoid.
Charged particles circumnavigating the brane undergo a phase shift of $2\pi$.

\subsection{D=5}

\begin{figure}[t]  
\vspace{4.5cm}  
\includegraphics{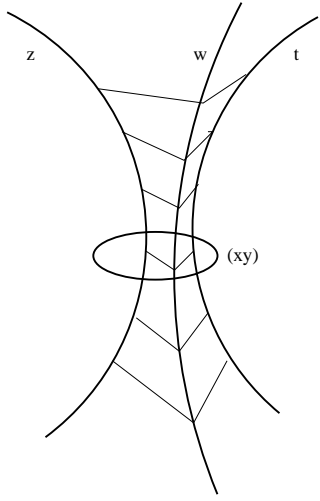}  
\vspace{2.5cm}  
\caption[]{
\small 
\addtolength{\baselineskip}{-.3\baselineskip}  
Construction of Dirac brane in $D=5$. The $(zwt)$ hypersurface is encircled
by the $(xy)$ current loop. The Aharonov-Bohm phase of an electron in the $(xy)$ loop
is quantized to $2\pi$, which quantizes the flux $F^*_{zwt}$ on the brane.}  
\label{dirac2}  
\end{figure}

Now consider the CS term in
$D=5$:  $\kappa \;\epsilon_{ABCDE}A^A\partial^B A^C \partial^D A^E$, which we note
is third order in the vector potential $A^A$.  Recalling that the CS
term induces an electric charge for the solenoid in $D=3$,
we are led to consider a field configuration that is second order in $A^A$ in
$D=5$. We thus consider the intersection of two 
Dirac branes in $D=5$.  Let us view the $3$-dimensional hypersurface dual to the
 $(xy)$ current plaquettes, carrying the 
 flux $ F^*_{zwt}$, as a {\em coherent
state} of photons. We denote this state by $\ket{\Phi_{xy}}$. 

 Recall the construction of coherent states: Let $\phi(x)$ be a quantum field
with canonical momentum, $\pi(x)$, and let $\phi_c$ be a classical background field.
Then the coherent state on a spacelike hypersurface is defined as:
\beq
\ket{\phi_c} = \exp\left( i \int \pi \phi_c \right) \ket{0}
\qquad
\makebox{whence,}
\qquad
\bra{\phi_c}\phi \ket{\phi_c} = \phi_c
\eeq
Note that if we have two classical configurations, $\phi_c^1$ and $\phi_c^2$
superimposed, then:
\beq
\ket{\phi^1_c\phi^2_c} = \exp\left( i \int \pi (\phi^1_c + \phi^2_c)\right)\ket{0} 
\eeq
hence,
\beq
\label{binom}
\bra{\phi_c}:\phi^2: \ket{\phi_c} = 2\phi^1_c\phi^2_c +(\phi^1_c)^2 + (\phi_c^2)^2
\eeq
In what follows we are
only interested in the cross-term for the superposition
of $N$ fields, since all other contributions will be zero by
the $\epsilon$-symbol. For $N$ superimposed configurations the cross-term coefficient
of $\phi^1_c \phi^2_c ...\phi^N_c$ is just $N!$.

Technically the coherent state of photons requires a Coulomb gauge 
for the vector potential, $\vec{A}$, which then has a well-defined canonical
momentum $\partial{\vec{A}}/\partial t$. This is always possible for our world-lines on a spacelike
hypersurface.
The field strength operator,
$F_{\mu\nu}$, then has a classical
expectation value in the coherent state,
on the brane which is a singular local form,
\beq
\label{cohere}
\bra{\Phi_{xy}} F_{\mu\nu} \ket{\Phi_{xy}} =
\frac{2\pi}{e}(g_{x\mu}g_{y\nu}-g_{y\mu}g_{x\nu})\delta(x)\delta(y)
\eeq
However, all that is relevant for our present considerations is that 
this expectation value satisfies eq.(\ref{fid}), so we
need deal only with exact integrals away from the singularity:
\beq
\label{cohere}
\int dx dy \bra{\Phi_{xy}}F_{xy} \ket{\Phi_{xy}} =
\frac{2\pi}{e}\;.
\eeq

We now form the intersection of two Dirac branes constructed
of $(xy)$ and $(zw)$ current plaquettes.
We describe this as a coherent state 
$\ket{\Phi_{xy}\Phi_{wz}}$.
We can again compute the expectation value of local operators
in this coherent state,
taking care to count the contractions of field operators
with photons in the state.
 We thus have:
\beq
\epsilon_{ABCDE}\int dxdydzdw\; \bra{\Phi_{xy}\Phi_{wz}}\partial^B
A^C\partial^{D}A^E \ket{\Phi_{xy}\Phi_{wz}} =
2\left(\frac{2\pi}{e}\right)^2 g_{At}
\eeq
The prefactor  of $2$ comes from the
two possible contractions of $dA$ with either $(xy)$ or $(wz)$
with the internal coherent fields \ie,  it is just the $2!$
in the coherent state of two superimposed classical 
fields as in eq.(\ref{binom}).  The exact orthogonality of the
surfaces
$(xy)$ and $(wz)$ does not affect this result, but the result
would be zero if the hypersurfaces were degenerate, \eg, $\ket{\Phi_{xy}\Phi_{xz}}$
yields a vanishing expectation value for the above
operator owing to the $\epsilon$ symbol in the CS term.

\begin{figure}[t]  
\vspace{4.5cm}  
\includegraphics{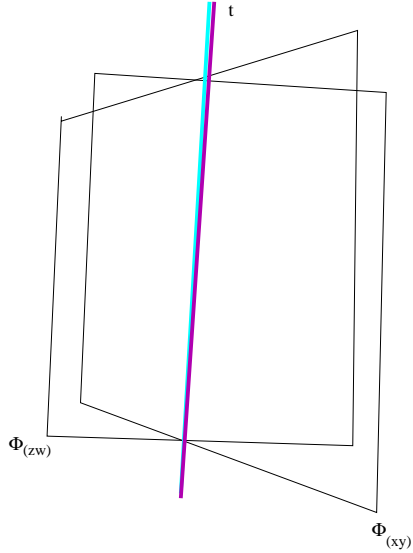}  
\vspace{3.5cm}  
\caption[]{
\small 
\addtolength{\baselineskip}{-.3\baselineskip}  
Intersection of Dirac branes in $D=5$. The Dirac brane carrying flux $\Phi_{xy}$
intersects $\Phi_{zw}$ to produce a charged worldline $t$. The charge is
determined by the CS term, $AdAdA$ and must equal the charge in
the Dirac quantization procedure.
}  
\label{dirac2}  
\end{figure}

Note that the intersection of the branes defines a world-line in the
$t$ direction.  It can thus be viewed as a particle world-line in $1+4$ dimensions.
If we now turn on the CS term, the 
intersecting brane world-line develops
an electric charge. We can compute the 
coupling of the world-line to an external coherent photon field
${A}^{ext}_\mu$  as:
\beq
\kappa\;\epsilon_{ABCDE}\int d^5x\; \bra{\Phi_{xy}\Phi_{wz}, {A}^{ext}_\mu} A^A\partial^B A^C\partial^D A^E 
\ket{\Phi_{xy}\Phi_{wz}, {A}^{ext}_\mu} =
3\times 2 \kappa\; \left(\frac{2\pi}{e}\right)^2 \int {A}^{ext}_0\; dt
\eeq
here the extra factor of $3$ counts the number of contractions with 
the external radiated photon field ${A}^{ext}$ by the vector potentials in the CS term,
where we have integrated by parts to remove the external photon momentum. 

We now enforce the self-consistency condition.
We demand that the induced world-line electric charge have
the same value as the original
defining electric charge, $e$, used to quantize
the brane flux. We thus have:
\beq
e \equiv 3\times 2 \kappa  \left(\frac{2\pi}{e}\right)^2; \qquad
\makebox{hence,} \qquad
\kappa = \frac{e^3}{24\pi^2}
\eeq
We have thus determined the CS  term coefficient $\kappa$. 
It agrees with the consistent anomaly coefficient
obtained from pure Weyl spinor triangle diagrams in $D=4$,
or equivalently Bardeen's LR symmetric form of the anomaly,
 \cite{Bardeen,framp,cth3,cth4}.

Note also that the intersection of the world-line by an $(xyzw)$
spacelike hypersurface is an instanton, \ie, 
it produces a nonzero value
of the matrix element in the intersecting brane coherent state,
$\int d^4x \;\bra{\Phi_{xy}\Phi_{wz}}F_{\mu\nu} F^{*\mu\nu}\ket{\Phi_{xy}\Phi_{wz}} = 16\pi^2/e^2$
where $F^*_{\mu\nu} = \half \epsilon_{\mu\nu\rho\sigma}F^{\rho\sigma}$
and $F=2dA$. 
The intersecting brane configuration
mimics features of the solenoid in $D=3$.

\subsection{Generalization to any $D$}

The construction we have described  generalizes 
readily to any odd $D$.  Current loops
in the $(xy)$ plane are stacked to generate a $D-2$
hypersurface carrying the flux dual to $F_{xy}$.
We again define this to be the coherent state $\ket{\Phi_{xy}}$
with the identical result of eq.(\ref{cohere}),
satisfying eq.(\ref{fid}). 

We now build the intersection of $(D-1)/2$ Dirac branes.
This is described
by the coherent state:
\beq
\ket{\Phi_{xy}\Phi_{uv}...\Phi_{zw}}
\eeq
This contains $(D-1)/2$, 2-forms, $F_{ij}$ and is dual to a $D-1$ dimensional
object, \ie, it can be taken to be a time-like world-line.
We compute the matrix element for a photon emission
in the presence of the CS term:
\bea
&& \kappa\;\epsilon_{ABC...DE}\int d^5x\; \bra{\Phi_{xy}...\Phi_{zw}, {A}^{ext}_\mu }
A^A\partial^B A^C ... \partial^D A^E 
\ket{\Phi_{xy}...\Phi_{zw}, {A}^{ext}_\mu}
\nonumber \\
&&
=\left[((D+1)/{2})!\right] \kappa \;  \left(\frac{2\pi}{e}\right)^{(D-1)/2} \int\; A_0
\; dx^0
\eea
The prefactor $((D+1)/{2})!$ is the generalization of the $3!$ in the $D=5$ case
counting the $((D-1)/{2})!$ contractions with the brane intersection coherent state, and 
$((D+1)/{2})$ contractions with the external photon.
The factor $ \left(\frac{2\pi}{e}\right)^{(D-1)/2}$ is just the $(D-1)/2$ brane flux factors of $2\pi/e$.  Again, though we chose orthogonal branes for simplicity,
the result applies to any  nondegenerate intersection of $(D-1)/2$ branes.

Demanding the consistency condition, \ie, 
that induced electric charge for the intersection
be equal to the defining Dirac charge $e$,
determines the CS term coefficient:
\beq
\kappa = \frac{e^{(D+1)/2}}{\left[((D+1)/{2})!\right] (2\pi)^{(D-1)/2}}
\qquad (D\;\;\makebox{odd})
\eeq
Stripping off a factor of $e$ (by convention)
this yields the consistent anomaly coefficient, $\kappa'$, for a right-handed Weyl spinor in even $\hat{D} = D-1$, 
\beq
\label{cac}
\kappa' = \frac{e^{\hat{D}/2}}{\left[((\hat{D}+2)/{2})!\right] (2\pi)^{(\hat{D})/2}}\qquad (\hat{D}\;\;\makebox{even}).
\eeq

\subsection{Covariant Anomaly Coefficients}

In $\hat{D}$-even space-time we consider Weyl spinor
theories $\bar{\psi}(i\slash{\partial} - A_L)\psi_L$
( $\bar{\psi}(i\slash{\partial} - A_R)\psi_R$ ) with currents $j_{\mu L} =  
\bar{\psi}\gamma_\mu \psi_L$ ( $j_{\mu R} =  
\bar{\psi}\gamma_\mu \psi_R$ ), and we
have the consistent anomalies:
\bea
\partial_\mu j^\mu_R & = & \kappa' \; dA_R ... dA_R 
\qquad
\partial_\mu j^\mu_L = -\kappa' \; dA_L ... dA_L
\eea
Now define:
\beq
A_L = V-A \qquad A_R = V+ A
\eeq
whence:
\bea
\partial_\mu j^\mu_R & = & \kappa' \;( dV ... dV +\frac{\hat{D}}{2}dA dV ... dV + ... + dA ... dA )
\nonumber \\ 
\partial_\mu j^\mu_L & = & -\kappa'  \;( dV ... dV -\frac{\hat{D}}{2}dA dV ... dV + ... 
+ (-)^{\hat{D}/2}dA ... dA )
\eea
or, in terms of vector and axial vector currents
($j= j_L+j_R$, $j^5=j_R-j_L$) we have:
\bea
\label{curva}
\partial_\mu j^\mu & = & \hat{D}\kappa' \;(dA dV ... dV + ...  )
\qquad 
\partial_\mu j^\mu_5  =  2\kappa'  \;( dV ... dV + ...  )
\eea
We see that the consistent anomalies violate both $j$ and $j^5$ conservation.
In the application to vector-like theories, such as QED, it is desireable to treat $V$
as a fundamental gauge field coupled to the vector
current, $j$, and to maintain
the conservation of $j$ in the presence of both $V$ and $A$. 
Thus, we can generate current correlators of $j$ and $j^5$ in
which $j$ is always conserved, by introducing counterterms, which is how
of Adler's original analysis arrives at the
covariant anomaly \cite{Adler}. 

Many counterterms are possible in higher $\hat{D}$. It is sufficient,
to determine the covariant anomaly coefficient,
to consider the leading terms on the rhs of eq.(\ref{curva}). Consider 
\beq
{\cal{O}} = -f\; \int\; A V dV ... dV
\eeq
If ${\cal{O}}$ is added to the action,  the currents are modified
by corrections:
\bea
\delta j & = & -\frac{\delta}{\delta V} {\cal{O}}
= -\half  f\; \left( \hat{D} A dV ... dV - ({\hat{D}}-{2} ) VdA dV .. dV  \right) 
\nonumber \\ 
\delta j_5 & = & -\frac{\delta}{\delta A} {\cal{O}}
= f\;V dV ... dV
\eea
whence the full currents now satisfy:
\bea
\partial_\mu (j + \delta j)^\mu  & = & \left(\hat{D}\kappa'  - {f} \right)dA dV ..dV + ... 
\qquad \nonumber \\
\partial_\mu (j_5 + \delta j_5)^\mu  & = & 2\left(\kappa' + \frac{f}{2}   \right)\;( dV ... dV 
) + ...
\eea
We thus specify $f$ by
demanding that the vector current is conserved, whence $f = \kappa' \hat{D}$:
\bea
\partial_\mu (j + \delta j)^\mu & = & 0
\qquad 
\partial_\mu (j_5 + \delta j_5)^\mu   =  2\kappa' \left(1+ \frac{\hat{D}}{2}   \right)\;( dV ... dV
+  ... 
)
\eea
Hence, using eq.(\ref{cac}) for $\kappa'$, the covariant anomaly coefficient is:
\beq
\tilde{\kappa } = 2\kappa \left(1+ \frac{\hat{D}}{2}   \right) = \frac{2e^{\hat{D}/2}}{(2\pi)^{{\hat{D}}/2} ({\hat{D}}/2)!}
\eeq
and the covariant axial current divergence is:
\beq
\partial_\mu \tilde{j}_5^\mu   =  \tilde{\kappa }\; (dV ... dV + ... )
\eeq
As a check, we see that this reproduces Adler's original result:
\beq
\partial_\mu \tilde{j}_5^\mu   =  \frac{e^2}{4\pi^2} dVdV  =  \frac{e^2}{8\pi^2} F \tilde{F}
\eeq
where $F\tilde{F}= 2dVdV$, where $\tilde{F}_{\mu\nu} = (1/2)\epsilon_{\mu\nu\rho\sigma}F^{\rho\sigma}$.
Our result is consistent with Frampton and Kephart \cite{framp} 
({\em c.f.} their eqs.(5.16) and (5.17) for
$2^{\ell-1} X_\ell$; note their result is quoted in
momentum space and imbeds the operator matrix element;
once one determines the general $1/(2\pi)^{\hat{D}/2}(\hat{D}/2)!$ behavior the
result is determined from the $\hat{D} =2,4$ results).

\vskip .1in
\noindent
\section{ Conclusions and Discussion}
\vskip .1in

\subsection{Summary of Results}

We can always introduce an integer ``index'' $N$ into the
anomaly and CS term coefficients which counts the ``number of (spectator) colors'' for
fermion loops.
In odd $D$ the Chern-Simons term for a $U(1)$
gauge theory is:
\be
-\kappa \;\epsilon_{ABC...CD}A^A\partial^B A^C ...\partial^C A^D 
\ee
where:
\be
\kappa = \frac{Ne^{(D+1)/2}}{\left[(({D}+1)/{2})!\right] (2\pi)^{({D}-1))/2}}
\ee
In even $\hat{D}$ the right-handed Weyl spinor
current coupled to a $U(1)$ gauge field has
the anomaly:
\beq
\partial_A \bar{\psi}\gamma^A \psi_R
= \frac{Ne^{\hat{D}/2}}{\left[((\hat{D}+2)/{2})!\right] (2\pi)^{(\hat{D})/2}}
\;\epsilon_{AB...CD}\partial^A A^B ...\partial^C A^D 
\eeq
where $ \psi_R = (1+\gamma^{\hat{D}+1})\psi/2$ (and correspondingly for
$\bar{\psi}\gamma^A \psi_L$ with a minus sign). 

In even $\hat{D}$ the covariant anomaly
for the axial current
current coupled to a $U(1)$ gauge vector-field $V$ has
the anomaly:
\beq
\partial_A \bar{\psi}\gamma^A\gamma^{\hat{D}+1} \psi
= \frac{2e^{\hat{D}/2}}{\left[((\hat{D})/{2})!\right] (2\pi)^{(\hat{D})/2}}
\;\epsilon_{AB...CD}\partial^A V^B ...\partial^C V^D 
\eeq
(this has additional terms if there is also an axial
photon as in \cite{Bardeen}).

For Yang-Mills theories the CS term has the structure
in odd $D$:
$\kappa \Tr(AdA ... dA + ...)$ where $A = A_A^aT^a$.
It is the $D+1$-th component of a current, $K$, whose divergence
is $\partial K = \Tr(F\wedge F \wedge ... F)$, with $(D+1)/2 $ field strength factors $F$.
The normalization of $T^a$ is irrelevant
since it can be absorbed into the definition of $e$.
Note that for an extra dimensional theory the CS term
coefficient steps through $\kappa$ as we cross a brane
with $N$ fermion species, \ie, $\kappa_R - \kappa_L = \kappa$
When compactifying onto a circle, $S_1$, $\kappa$ is in principle
arbitrary because there is no net
anomaly on a boundary. 
This is the analogue of the $\theta$ term in $D=5$.
The $D=5$ nonabelian CS term under suitable compactification
becomes the Wess-Zumino-Witten term and the coefficient carries over
\cite{cth2}.

\subsection{Discussion and Summary}

Axial anomalies are traditionally viewed as intrinsically
fermionic in origin,  \cite{S,BJ,Adler,Bardeen,AdlerBardeen}. 
However,  axial anomalies are often present in purely bosonic effective theories
and their coefficients can be fixed modulo an integer by self-consistency arguments. 
A familiar example is a low energy effective lagrangian
of the $\pi^0$ and the photon, including a term of the form
$(N_c\alpha/2\pi f_\pi)\pi^0 F_{\mu\nu}\widetilde{F}^{\mu\nu}$.
This is a term in the overall
Wess-Zumino-Witten term \cite{Wess,Witten} of 
low energy QCD chiral dynamics. Its coefficient
can be determined by the underlying anomalous quark loop
structure \cite{BFGM}, and the WZW term then gives a complete bosonic description of the anomaly structure. 

However, we can determine the WZW term coefficient, hence the anomaly coefficient,
from a purely bosonic argument, without resort
to fermion loops. The full WZW term is a necessary 
part of the effective action in
the IR theory required to generate a complete physical description of 
the skyrmion, the low energy description of the baryon \cite{Witten}. 
The skyrmion is topologically stable 
and has conserved topological currents, \eg, 
 the singlet Goldstone-Wilczek  \cite{GW} current. 
Noether variation of the WZW term {\em wrt} the $\omega$ meson generates the
gauge invariant form of the Goldstone-Wilczek current (this requires 
care in the Standard Model \cite{HHH}). 
We can therefore determine the coefficient of the WZW term 
by choosing the Goldstone-Wilczek
charge, the baryon number of the skyrmion, to be an integer.  The WZW term also
has been argued to control the spin and statistics
of the skyrmion, confirming its interpretation
as the low energy description of the baryon in QCD when the index
$N\equiv N_c=3$ \cite{Witten,Witten2}. Remarkably,
this determines the anomaly
coefficient and $\pi^0$ decay without resort to a fermion loop 
calculation. 

We can determine the CS term coefficient
for a nonabelian Yang-Mills theory in $D=5$ in a similar fashion. In fact, 
a $D=5$ Yang-Mills theory of flavor, suitably compactified, becomes
a chiral lagrangian of pions (the $A^5$ zero modes) 
coupled to flavored gauge fields in $\hat{D}=4$,
and the topological aspects of QCD emerge
from the $D=5$ CS term \cite{cth2}. 
In the $D=5$ Yang-Mills theory there exists an  ``instantonic vortex,"
which is
a stable soliton solution, and under compactification to $\hat{D}=4$ this object
holographically maps onto the skyrmion, \cite{atiyah};  
naturally, the $D=5$ CS currents, in turn, match onto the skyrme
currents,  \cite{cth}.  Again, demanding that
the instantonic vortex carry an integer charge dictates
the CS term coefficient, hence the ``consistent anomaly''
coefficient in $\hat{D}=4$.
Via fermion loops, this involves 
the computation of triangle and box diagrams
 \cite{Bardeen}, but the form of the consistent anomaly
can be immediately obtained from the boundary variation of the 
properly normalized CS term
under a gauge transformation
\cite{jackiw2}.  

In the present paper we have illustrated a general
construction of a field configuration that carries the $U(1)$ CS charge 
in any $U(1)$ theory in any $D$-odd dimension. 
This begins with Dirac's solenoid construction of the magnetic 
monopole in $D=3$ for a $U(1)$ gauge theory.  
The solenoidal flux is chosen so that the
phase shift of an encircling ``electron'' of charge $e$  
is  $2\pi $, thus ``cloaking'' the solenoid. This
quantizes the monopole's magnetic charge as $g_m = \hbar/2e$.

However, when we turn on the CS term, $\kappa AdA = \kappa \epsilon_{\mu\nu\rho}A^\mu\partial^\nu A^\rho$, three things happen: (1) the theory
becomes that of a massive photon \cite{schonfeld, deser} and we can easily see that
monopole solutions {\em cannot exist} when $\kappa \neq 0$;
(2) the solenoid must therefore be either
a closed loop, or an infinite line in $D=1+2$;
(3) with $\kappa \neq 0$, the Dirac solenoid becomes physical: an
electric current flows in a closed solenoid loop;  alternatively,
in $D=1+2$, an infinite timelike solenoid becomes
the world-line of a charged particle that can emit or absorb photons.
The Dirac solenoid has become a bosonic string 
with $\kappa \neq 0$.   

We must then impose 
a ``consistency condition," \ie, that the induced charge carrier in the 
solenoid current loop 
(\ie, the electric charge of the $1+2$ particle solenoid world-line)
have the same value as the defining electric
charge, $e$, employed in the original Dirac quantization 
condition.
With the consistency condition 
the coefficient of the CS term is determined and found to
be $e^2/4\pi$.
This, in turn, dictates the axial anomaly coefficient in $\hat{D}=2$. 
  
We generalize the solenoid construction to Dirac branes,
first introduced by Teitelboim \cite{teit}.
In  higher odd dimensionality $D$, we consider
Dirac branes that are extended objects that can always
be encircled by a charged particle 
world-line loop. The branes are therefore $D-2$ hypersurfaces. 
With the encircling loop we can impose the Dirac flux
quantization condition on a Dirac brane, analogous to the solenoid
construction in $D=3$. In \eg, $D=5$, this becomes a three dimensional extended surface
carrying a flux $F^*_{\mu\nu\rho}$.

The simultaneous intersection of $(D-1)/2$ of
Dirac branes defines a world-line in $D$ dimensions.
This intersection  
becomes the world-line of an electrically charged particle
when the $D$ dimensional CS term, $AdAdA...dA$, is switched on.
We must impose the consistency
condition that this 
world-line carry the defining
charge used to quantize the Dirac brane flux. 
In this manner we thus obtain the CS term coefficient in any odd $D$.

Compactification of the $D$-odd theory to an even $\hat{D} = D-1$ 
theory generates the consistent anomaly
coefficient on the boundary branes from the CS term coefficient.
One can introduce fermions on the $\hat{D}$ boundaries to cancel these
anomalies if one wants an anomaly free theory \cite{cth3}.
In this picture, fermions are ``spectators'' to the bosonic theory that can
remedy the uncancelled bosonic anomalies.  This is a
``yin-yan" view of the role of bosons vs. fermions
in axial anomalies.  

Consistent anomalies, equivalent
to  ``left-right symmetric anomalies'' \cite{Bardeen}, 
occur in both the vector and axial-vector currents,
are generated by Weyl spinor loops. 
For covariant anomalies
the gauged currents (\eg, vector current in QED) are conserved for any background fields
upon including counterterms. Adler's
result for QED was a covariant anomaly (see section 3). In Yang-Mills theories, covariant
anomalies are gauge invariant, while consistent anomalies are not. 
To construct the covariant anomaly in vectorlike theories, in which
the gauged vector currents are conserved, requires  
including the ``Bardeen counterterm'' 
into the action \cite{Bardeen}. We can easily
construct the counterterm in any even $\hat{D} = D-1$
and demand vector current conservation, to obtain the covariant anomaly coefficient.
We thus arrive at a final result for the consistent and covariant anomaly coefficients in any even $\hat{D}$.  Our results confirm the fermion loop
calculations of Frampton and Kephart in any $\hat{D}$ \cite{framp}.
\footnote{
Note that there are important instances in which we do not
want to enforce vector current
conservation with the Bardeen counterterm, such as the case of the Standard Model
in which the gauged left-handed currents must be anomaly free. This
requires the Standard Model counterterm of ref.\cite{HHH} 
and leads to some interesting modifications in the low energy effective
theory via the WZW term. }

Our main result, in summary, is 
that there always exists, by construction, a
gauge field configuration (albeit singular) 
that inherits an electric charge in the presence of 
the CS term in any odd $D$, for any gauge theory,
since any gauge group $G$ contains $U(1)$.
This field configuration can be taken to be a collection
of $(D-1)/2$ intersecting Dirac branes in a $U(1)$
submanifold of $G$. The field configuration
on a spacelike hypersurface, 
cutting through the brane intersection is an instanton with
unit Pontryagin index. Hence,
this brane intersection is related to the notion of an instantonic vortex. 
The consistency of the induced electric charge on the intersection
with the Dirac quantization condition determines the CS term
coefficient and hence the consistent anomaly. The covariant anomaly
follows from the judicious choice of counterterm.

Axial anomalies can  be viewed as purely
bosonic in origin.
They are intrinsically topological and arise holographically
from  Chern-Simons terms in odd $D$. We've
shown that axial anomalies trace directly from the Dirac monopole construction.
A Chern-Simons term destroys monopoles, and thus destroys the duality of
electrodynamics, essentially enforcing a vector potential description
rather than allowing, \eg, an axial vector potential. This leads to the violation
of axial current conservation.

\vskip .1in
\noindent
{\bf Acknowledgements}
\vskip .1in
We thank Bill Bardeen, James Bjorken, Paul Frampton, Jeff Harvey, Richard Hill, 
Tom Kephart and Cosmas Zachos 
for helpful discussions.

\newpage    
\setcounter{equation}{0}
\renewcommand{\theequation}{A.\arabic{equation}}      
\subsection{Weyl Spinor Anomaly in $\hat{D}=2$}    

\begin{figure}[t]  
\vspace{3.0cm}  
\includegraphics{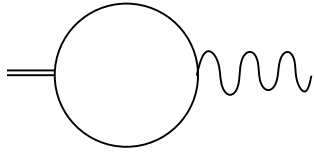}  
\vspace{2.5cm}  
\caption[]{
\small 
\addtolength{\baselineskip}{-.3\baselineskip}  
Anomaly in $\hat{D}=2$.
}  
\label{dirac2}  
\end{figure}

We can readily demonstrate the consistent anomaly in the theory for a single Weyl fermion
coupled to a photon in $\hat{D}=2$,  
\be
S = \int d^2x\; \bar{\psi}(i\slash{\partial} - \slash{A}_L)\psi_L
\ee
where, $\psi_L = (1-\gamma^3)\psi/2$.
A Euclidean calculation suffices. 
We choose the $\gamma$-matrices
to be the Pauli matrics, 
$\gamma_0 = \tau^x$, $\gamma_1 = \tau^y$, $\gamma^3 = \tau^z$.
Then, $\slash{p} =p_0\tau^x+p_1\tau^y$,
$\Tr(\tau^z\slash{a}\slash{b}) = 2i\epsilon_{ij}a^ib^j$.  
The divergence of the
current $j_{\mu L} = \bar{\psi}\gamma_\mu \psi_L$ has a matrix
element to a single photon 
of polarization $\epsilon_\mu$ given by:
\bea
\bra{0}\partial_\mu j_L^\mu\ket{\epsilon} & = & -\int \frac{d^2\ell}{(2\pi)^2}\Tr \left[\frac{\slash{\ell}\slash{\epsilon}(\slash{\ell} +\slash{q}) \slash{q}(1-\tau^z)
 }{2(\ell+q)^2\ell^2}\right]
\nonumber \\
&= &  \int_0^1 \;dx \; {q}^2x(1-x) \int \frac{d^2\bar{\ell}}{(2\pi)^2}\Tr
\left[ \frac{\tau^z
\slash{\epsilon}\slash{q} }{2(\bar{\ell}^2+x(1-x)q^2)^2}\right]
\nonumber \\
& = & -\frac{1}{4\pi}\epsilon_{\mu\nu}q^\mu \epsilon^\nu 
\eea
where $\bar{\ell} = \ell + xq$, and
the divergent part of the  $\bar{\ell}$
integral is zero owing to the identity
$\gamma_\mu \slash{\ell} \gamma^\mu = 0$. 
This is the ``consistent'' anomaly:
\be
\partial^\mu \bar{\psi}\gamma_\mu \psi_L = -\frac{1}{4\pi}\epsilon_{\mu\nu}\partial^\mu A_L^\nu .
\ee


\end{document}